\documentclass[10pt,aps,prl,twocolumn,superscriptaddress]{revtex4}

\usepackage{graphicx}
\usepackage{amsmath,amssymb}
\usepackage{color}
\usepackage{url}
\usepackage{xfrac}
\usepackage{upgreek}
\usepackage[outdir=./]{epstopdf}
\usepackage{siunitx}

\graphicspath{{./Images/}}

\usepackage[linktocpage=true, colorlinks=true, urlcolor=blue, linkcolor=blue, citecolor=blue]{hyperref}
\usepackage{cleveref}

\begin{document}

\title{Spatially or temporally resolved phase-based optoretinography with actively-tracked point-scanning optical coherence tomography}

\author{L\'eo Puyo}
\affiliation{Medical Laser Center L\"ubeck GmbH, Peter-Monnik-Weg 4, 23562 L\"ubeck, Germany}
\affiliation{Institute of Biomedical Optics, University of L\"ubeck, Peter-Monnik-Weg 4, 23562 L\"ubeck, Germany}
\affiliation{gl.puyo@gmail.com}

\author{Jonas Franke}
\affiliation{Medical Laser Center L\"ubeck GmbH, Peter-Monnik-Weg 4, 23562 L\"ubeck, Germany}
\affiliation{Institute of Biomedical Optics, University of L\"ubeck, Peter-Monnik-Weg 4, 23562 L\"ubeck, Germany}

\author{Felix Zeisberger}
\affiliation{Medical Laser Center L\"ubeck GmbH, Peter-Monnik-Weg 4, 23562 L\"ubeck, Germany}
\affiliation{Institute of Biomedical Optics, University of L\"ubeck, Peter-Monnik-Weg 4, 23562 L\"ubeck, Germany}

\author{Roland Rocholz}
\affiliation{Heidelberg Engineering GmbH, Max-Jarecki-Str. 8, 69115 Heidelberg, Germany}

\author{Gereon H\"uttmann}
\affiliation{Institute of Biomedical Optics, University of L\"ubeck, Peter-Monnik-Weg 4, 23562 L\"ubeck, Germany}
\affiliation{Medical Laser Center L\"ubeck GmbH, Peter-Monnik-Weg 4, 23562 L\"ubeck, Germany}
\affiliation{Airway Research Center North (ARCN), Member of the German Center for Lung Research (DZL)}

%\date{\today}

\begin{abstract}
Optoretinography is a promising approach to measure visual function but its clinical translation has remained confined to research settings due to demanding hardware requirements. Here, we demonstrate that high-quality phase-based optoretinography can be performed using conventional point-scanning spectral-domain optical coherence tomography. We show that the need for high imaging speed can be circumvented by accurate sampling of phase changes at consistent spatial location. Light-evoked outer-segment expansion could be temporally resolved in B-scan series acquired at a fixed position when filtering off-location B-scans, and spatially resolved in volumes acquired at rates as low as 1~Hz. These results indicate that the technical barrier for phase-based optoretinography is substantially lower than previously assumed and pave the way toward broad clinical translation.
\end{abstract}

\maketitle

Optoretinography encompasses noninvasive optical methods probing the function of retinal neurons~\cite{jonnal2021, matteson2025}. It aims at detecting changes in reflectivity, optical path length, or scattering properties associated with phototransduction. Optoretinography holds strong potential for quantitative characterization of retinal function with foreseeable applications in visual field assessment and in the development of vision-restoring therapies~\cite{chen2025recent}.
% first demonstration too complex for clinic
Although optoretinography was first explored with intensity-based approaches~\cite{bizheva2006, jonnal2007, grieve2008intrinsic}, phase-sensitive OCT arguably elevated the technique to a new level by enabling quantitative measurement of outer-segment (OS) deformation with almost nanometer precision~\cite{Hillmann2016PNAS}. Pioneering work in phase-based optoretinography was performed at high spatiotemporal resolution with OCT combined to digital or hardware adaptive optics~\cite{Hillmann2016PNAS, zhang2019cone, pandiyan2020}. Functional readout from individual photoreceptors was demonstrated~\cite{zhang2019cone}, providing detailed insight into cellular responses under various temporal~\cite{Pfaeffle2022PhaseSensitive, tomczewski2025photopic}, and spectroscopic stimulation conditions~\cite{zhang2019cone}. However, despite its high relevance for basic research, optoretinography at cellular-resolution remains impractical for widespread clinical deployment. The required hardware, typically including deformable mirrors, wavefront sensors, and/or high-speed cameras, adds cost and complexity, preventing clinical adoption.
%lower spatial resolution is more relevant
Moreover, cellular resolution seems unnecessary for clinical usage, where the spatial scale of interest exceeds that of individual cells. Lower spatial resolution would, in fact, facilitate clinical translation by enabling imaging over a wider field of view. Coarse-scale optoretinography has shown excellent agreement with adaptive-optics-assisted measurements~\cite{jiang2022coarse}, alleviating concerns about the validity of optoretinography signals obtained at lower spatial resolution.

%temporal resolution
High temporal resolution may similarly not be essential for clinical use as the currently clinically established biomarker for phase-based optoretinography is the maximum OS extension~\cite{Lassoued2021, wendel2024, liu2025longitudinal}. Intensity-based approaches also found that the amplitude of optoretinograpy response appears as the key biomarker~\cite{xu2024, gaffney2024, wongchaisuwat2024}.
The initial contraction within the first milliseconds of stimulation and the detailed shape of the deformation are of high interest for basic research but have not yet demonstrated clinical value.
High imaging speed is nevertheless beneficial for mitigating eye movement, reducing motion-induced distortions in scanned images, and facilitating the identification of phase fluctuations in the midst of global eye motion.
Temporal oversampling can be leveraged by the extended Knox-Thompson method to improve phase-based optoretinography by filtering phase fluctuations caused by speckle changes unrelated to the OS deformation~\cite{Spahr2019}. Phase-based optoretinography could be demonstrated with 100~Hz B-scan series combined to extended Knox-Thompson filtering in anesthetized mice~\cite{pijewska2021}. However, the translation of this approach to humans is challenging, as anesthesia substantially reduces eye motion, while involuntary eye motion represents a major challenge for measurements in awake, unanesthetized subjects.
% eye movement and phase errors
Instability of the eye during B-scan series leads to variations in sampling location perpendicular to the scanning axis that cannot be corrected in post-processing. The measured phase variations in the B-scan series ambiguously results from both stimulus-evoked temporal changes and spatial variations in OS length. Increasing the B-scan rate does not resolve this limitation.
Velocity-based optoretinography is another form of phase-based optoretinography, which monitors the temporal derivative of OS length over a sliding window of approximately 10~ms~\cite{Vienola2022}. However, it does not provide the total OS expansion and although reproducible results have been reported in humans, it assumes negligible eye motion within the sliding window, limiting its clinical robustness.
% our work
Here, we demonstrate that phase-based optoretinography can be performed in B-scan series using conventional point-scanning OCT by combining active eye tracking with post-processing rejection of mislocalized B-scans. We further show spatially resolved optoretinography in volume series acquired at $\sim$1~Hz with active tracking.

%%%%%%%%%%%%%%% METHODS %%%%%%%%%%%%%%%%%%%

%\section*{Methods}

We used a research-grade point-scanning spectral-domain Spectralis OCT (Heidelberg Engineering GmbH, Germany) operating at a \SI{250}{\kilo\hertz} A-scan rate with a superluminescent diode centered at $\lambda = 870~\mathrm{nm}$, providing an axial resolution of 6.9~\textmu m in tissue, sampled with an axial pixel size of 3.9~\textmu m.
%A superluminescent diode centered at $\lambda = 870~\mathrm{nm}$ with a bandwidth of \SI{50}{\nano\meter} provided an axial resolution of 6.9~\textmu m in tissue, sampled with an axial pixel size of 3.9~\textmu m.
The beam diameter at the pupil was \SI{1.7}{\milli\meter}, corresponding to a diffraction-limited lateral resolution of approximately 5~\textmu m~\cite{Spaide2022lateral}.
%tracking
The Spectralis employs a scanning laser ophthalmoscope (SLO) to track and compensate eye motion in real time. Active tracking repositions the OCT beam to the intended retinal location at the beginning of each B-scan and stores frames only if the beam position remains within predefined tolerance bounds, set to the lateral point-spread function for motion perpendicular to the scanning axis. Active tracking was used for both B-scan and volume series and tracking positions were saved together with the OCT raw data.
% stimulation
White-light photostimulation was delivered by a light-emitting diode spatially obstructed by a circular mask.
% OCT data processing
Raw OCT data were reconstructed using standard processing steps including $k$-space linearization and dispersion compensation. Axial motion was corrected by cross-correlation to a reference A-scan. Phase changes were evaluated between the inner segment/outer segment (IS/OS) junction and the cone OS tips (COST). Each layer was manually segmented and averaged over two axial pixels. The phase difference $\Delta \varphi$ was converted to physical displacement with $\Delta \mathrm{z}= \lambda \frac{\Delta \varphi}{4 \pi n}$, where $n$ denotes the refractive index and was approximated to 1.33.
% subjects
Investigations were done with dark-adapted healthy volunteers with a medically dilated pupil. Results from three subjects (22, 28, and 32 years old) are shown. Experimental procedures adhered to the tenets of the Declaration of Helsinki and informed consent was obtained from the subject.
%All experiments were performed in accordance with relevant guidelines and regulations.
Compliance with all relevant safety rules was confirmed by the responsible safety officer.

%\section*{Results}

\begin{figure}[t!]
\centering
%\includesvg[width=1\linewidth]{1_BscanRejection.svg}
\includegraphics[width = 1\linewidth]{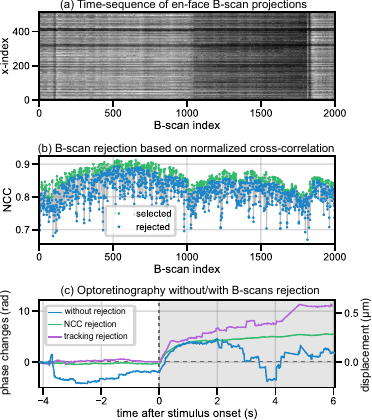}
\caption{Rejecting B-scans based on normalized cross-correlation (NCC) improves optoretinography measurements.
(a) En-face projections of B-scans reveal temporal irregularities.
(b) B-scans with lowest NCC to the averaged B-scan are rejected over a sliding window.
(c) Phase-based optoretinography is dramatically improved with this rejection scheme.
}
\label{1_BscanRejection}
\end{figure}

Series of 2000 B-scans consisting of 512 A-scans spanning over $10^\circ$ were acquired at a fixed location.
%(a)
Although B-scans were recorded with active tracking and registered along the fast scanning axis by cross-correlation, residual irregularities over time are seen on the en-face projections shown in Fig.~\ref{1_BscanRejection}(a). This is due to the residual out-of-plane motion despite active tracking, which cannot be corrected for in post-processing along the slow scanning axis.
%(b)
To mitigate this effect, we introduced a rejection step to exclude B-scans shifted perpendicularly to the scanning axis from the intended B-scan location. Assuming the scanned area slightly fluctuates around the correct position, the time-averaged B-scan should reflect the correct position, and larger distance from the targeted location should result in a higher dissimilarity of B-scans to this time-average. For each B-scan, the normalized cross-correlation (NCC) with the average of its 500 temporally closest neighbors was computed, shown in Fig.~\ref{1_BscanRejection}(b). Within sliding windows of 40 B-scans, the 70\% of B-scans with the lowest NCC values were discarded. This strategy ensured uniform rejection despite slow variations in NCC across the acquisition.
%(c)
As shown in Fig.~\ref{1_BscanRejection}(c), despite sparser sampling and a $\sim70\%$ reduction in temporal resolution, this rejection scheme markedly improves phase stability by avoiding phase errors that otherwise accumulate and corrupt the final OS extension value. The optoretinography signal resulting from continuous stimulation applied from onset until the end of the measurement is largely improved (the grey area indicates the stimulation period). The rejection ratio is chosen to meet a trade-off between improving spatial consistency (high ratio) and preserving temporal resolution (low ratio); the value of 70\% consistently gave stable results across all subjects and acquisitions in our measurements.
This scheme proves more efficient than a rejection based on residual position errors (with identical rejection criteria) provided by SLO tracking (purple curve). Variations in intensity seen in Fig.~\ref{1_BscanRejection}(a) and attributable to changes in aperture from eye movements do not appear to affect optoretinography when using the NCC-based rejection.

\begin{figure}[t!]
\centering
%\includesvg[width=1\linewidth]{2_BscanPower.svg}
\includegraphics[width = 1\linewidth]{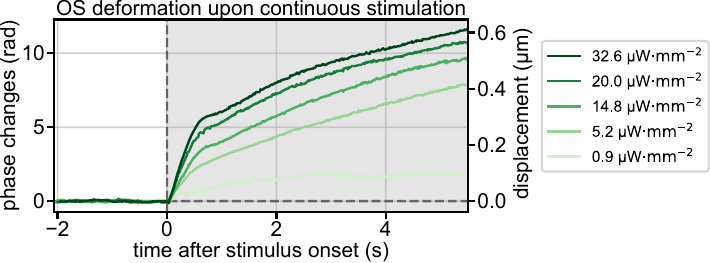}
\caption{Optoretinography with series of B-scans with continuous stimulation of different power. The phase is highly stabilized before stimulation onset, and the expected increase in OS deformation with higher stimulation power is verified.
}
\label{2_BscanPowerDuration}
\end{figure}

We next evaluated optoretinography responses for different stimulation powers using identical B-scan acquisition parameters (2000 B-scans, 512 A-scans over $10^\circ$) and  $70\%$  rejection criteria. The stimulation again consisted of continuous illumination from onset until the end of the measurement. As shown in Fig.~\ref{2_BscanPowerDuration}, before stimulation onset, phase variations remain close to zero, demonstrating high phase stability. After stimulation, the OS elongation is clearly resolved in all measurements, and the expected increased optoretinography response with stronger stimulation is verified.

Phase-based optoretinography was further assessed in volumetric acquisitions. Series of six volumes ($512 \times  200$ A-scans) covering $10^\circ \times 3^\circ$ were recorded with active tracking, resulting in an effective volumetric rate of approximately 1~Hz. To compensate for eye movement, an additional en-face registration along the fast-scanning axis was performed using a strip-based approach similar to that described by Chen et al.~\cite{chen2023}. The volume exhibiting minimal eye motion according to the SLO tracking data was selected as reference. Iteratively decreasing strip widths (100, 50, 25, and 5 pixels) were used to compute cross-correlations on en-face projections and integer-pixel shifts were applied.
% thereby avoiding interpolation of complex-valued data.
Voxel-wise phase differences between IS/OS and COST were smoothed in complex domain by convolution with a Gaussian filter of 5-pixel width.

\begin{figure}[]
\centering
%\includesvg[width=1\linewidth]{3_VolumesTimeChanges.svg}
\includegraphics[width = 1\linewidth]{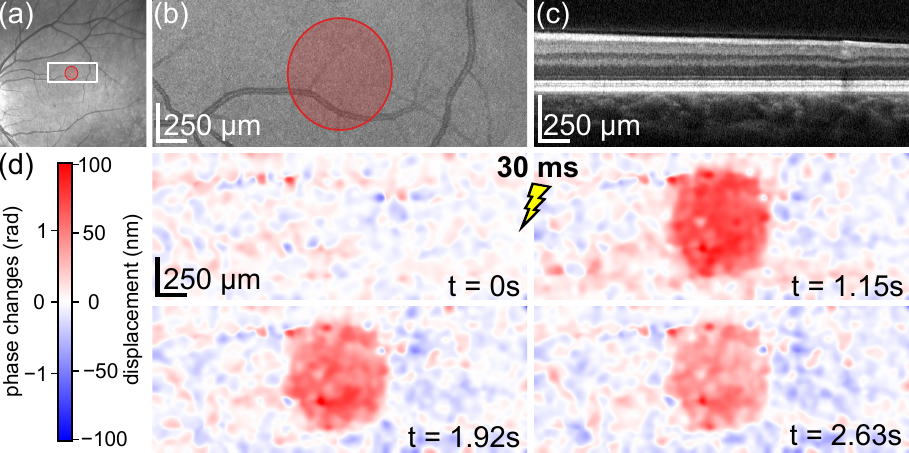}
\caption{Optoretinography in volume series acquired at $\sim1~$Hz with a 30~ms flash stimulation. (a) SLO showing the imaged region. (b-c) en-face and cross-sectional OCT view. (d) Optoretinograms showing the detected OS phase variations within the stimulated central region.
}
\label{3_VolumesTimeChanges}
\end{figure}

First, we used a 30~ms stimulation flash, delivered at the beginning of the third volume acquisition over a disk with an average power of 42~\textmu W$\cdot \mathrm{mm}^{-2}$. Figure~\ref{3_VolumesTimeChanges}(a-c) shows en-face and cross-sectional OCT images along the fast scanning axis, with the region of interest indicated on the SLO image. Optoretinograms below show the accumulated phase increase; the indicated time corresponds to the end of each volume acquisition. The local OS elongation in the first post-stimulus volume is visible as well as the reversal of the elongation, observed in the third and fourth optoretinograms. Local elongation is revealed with a resolution of a few tens of nm and the OS elongation can be measured even beneath retinal vessels. Spatial Gaussian smoothing results in resolution loss but effectively mitigates phase noise due to inconsistencies in sampling location by averaging out positive and negative phase errors relatively to the correct phase value. This smoothing is particularly necessary as no registration was performed along the slow scanning axis.

\begin{figure}[]
\centering
%\includesvg[width=1\linewidth]{4_VolumePower.svg}
\includegraphics[width = 1\linewidth]{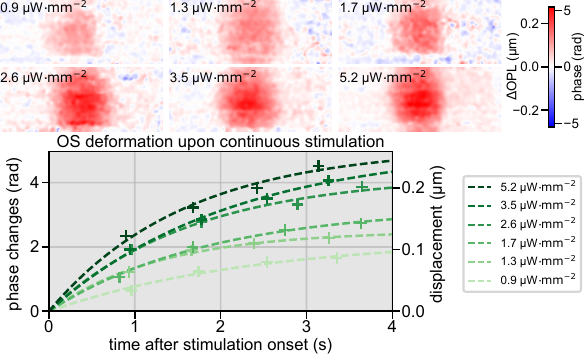}
\caption{Phase-based optoretinography in volume series with photostimulation over a central circular area at different powers. Top: en-face optoretinograms showing accumulated phase. Bottom: corresponding temporal phase traces.
}
\label{4_VolumePower}
\end{figure}

Finally, optoretinography was carried out in volume series for different stimulation powers using the same volume acquisition parameters as before. Continuous illumination was applied within a central circular region from the beginning of the second volume acquisition until the end of the measurement.
Accumulated phase maps shown in Fig.~\ref{4_VolumePower} were obtained after spatiotemporal Gaussian smoothing (5 pixels laterally, 1 pixel temporally). Temporal phase traces were calculated by spatial averaging phases within a central circular region of interest of 100 pixels, centered on the detected maximum phase changes.
As with B-scan series, higher stimulation powers produced stronger optoretinography responses. Because of the volumetric rate of approximately 1~Hz, stimulation must be limited to provoke an OS deformation resulting in a phase increase smaller than $\sim \pi /$volume. This results in a more severe trade-off between low signal-to-noise ratio at low power and phase wrapping at high power. Nevertheless, volumetric measurements prove comparatively robust to occasional misaligned B-scans, as residual phase errors are partially averaged out by spatial smoothing.
%Moreover, measurements over several volumes increase the separation between OS elongation responses for different stimulus intensities, partially compensating for the low stimulation levels.

%%%%%%%%% DISCUSSION %%%%%%%%%%

%\section*{Discussion}

Phase-based optoretinography was originally demonstrated using full-field swept-source OCT~\cite{Hillmann2016PNAS}, which offers the possibility to register volumes series with high precision and provides intrinsic lateral phase stability, necessary for digital refocusing~\cite{Hillmann2016Aberration}. However, phase-based optoretinography only requires the relative phase difference between the IS/OS junction and COST within the same A-scan. Relative phases within individual A-scans of spectral-domain OCT exhibit excellent stability, probably exceeding that of full-field swept-source OCT due to reduced multiple scattering and shadowing artifacts below blood vessels.
The improvement of phase measurements after rejection of B-scans highlights the importance of following phase evolution at consistent spatial locations. The use of active tracking to perform in vivo phase-sensitive measurements with point-scanning OCT had already been reported for measuring blood flow-induced tissue motion~\cite{desissaire2021}, and by Wong et al. for high-resolution optoretinography~\cite{wong2025}. In the latter case, adaptive optics was used and phase variations were sampled in volumes acquired at 20~Hz, resulting in a field of view of $0.05^\circ \times 0.5^\circ$. Interestingly, Wong et al. reported seemingly stable optoretinography without additional B-scan rejection as required in our case. This may reflect differences in tracking performance, but we also hypothesize that our larger spatial scale may introduce additional challenges. High-resolution SLO provides richer texture and reliable landmarks for tracking~\cite{sheehy2012high}, whereas these features are less distinct at coarse resolution. Moreover, tracking errors on the order of the point-spread function remain comparable to the photoreceptor size at high resolution, helping preserve consistent phase differences between IS/OS and COST. At coarser scales, however, topographic variations in OS length within the lateral resolution become more pronounced, leading to stronger phase errors when tracking inaccuracies occur.

Simple cross-correlation methods were used for lateral and axial registration. Residual registration errors are visible in en-face images, especially as no numerical correction was applied along the slow-scanning axis. Gaussian smoothing was key in mitigating the detrimental effects of approximate registration by averaging phase errors, which is a key advantage of phase-based optoretinography.

%improvement
Several directions exist to further improve robustness and measurement quality, including better real-time tracking and volume registration and segmentation in post-processing. The extended Knox-Thompson method can be expected to further improve robustness against phase errors in B-scan series~\cite{Spahr2019, pijewska2021}. One limitation of volumetric optoretinography is the trade-off between stimulation power and phase wrapping. Stronger stimulation would improve signal quality but also increase the risk of phase wrapping when volumes are acquired at slow rates.
Active tracking cannot compensate for fast eye movement such as saccades. This makes volumes acquisition time dependent on fixation performance, resulting in longer acquisition time which could lead to critical phase wrapping in subjects with unstable fixation. Higher A-scan rates would reduce the amplitude of eye movement during B-scans and could be used to image a larger field of view or to prevent phase wrapping by enabling faster volumetric acquisition.
%The effective A-scan rate was approximately half of the normal one when using active tracking. 
% SSADOR
Our approach can be compared to intensity-based split-spectrum amplitude-decorrelation optoretinography~\cite{chen2023, wongchaisuwat2024}, in that both enable optoretinography measurements using conventional, commercially available point-scanning OCT systems. The phase-based approach however offers quantitative measurements of the OS extension, which is the biomarker with currently established clinical relevance. Furthermore, we speculate that the phase-based approach may also be more robust to lateral registration imprecision due to the spatial averaging of phase errors.

%%%%%%%%% CONCLUSION %%%%%%%%%%

%\section*{Conclusion}

In summary, we demonstrated spatially or temporally resolved phase-based optoretinography from B-scans and volume series acquired with conventional spectral-domain point-scanning OCT.
Combining active tracking with a rejection of off-location B-scans enabled reliable optoretinography measurements in B-scan series over extended duration despite fixational eye movement.
Phase changes could be also quantitatively followed in volume series over a $10^\circ \times 3^\circ$ field of view at volumetric rates as low as $\sim1~$Hz. This approach holds strong potential for further development by improving tracking precision, registration and segmentation, and using higher A-scan rates. This work was achieved with a commercially available OCT system showing that the technical barrier for phase-based optoretinography is substantially lower than previously assumed. Although repeatability studies are missing and results are presented from only 3 young subjects, the approach looks promising for the clinical translation of optoretinography.

\section*{Funding}
Bundesministerium f\"ur Bildung und Forschung (BMBF 13N15432).

\section*{Disclosures}
LP: Heidelberg Engineering GmbH (F),
RR: Heidelberg Engineering GmbH (E),
GH: Heidelberg Engineering GmbH (C, P, F).

\bibliographystyle{unsrt}
\bibliography{./Bibliography}

\end{document}